\documentclass[]{interact}

\usepackage[caption=false]{subfig}

\usepackage[numbers,sort&compress]{natbib}
\usepackage{amsmath,cleveref,subfig,graphicx,tikz,verbatim}

\bibpunct[, ]{[}{]}{,}{n}{,}{,}
\renewcommand\bibfont{\fontsize{10}{12}\selectfont}
\makeatletter
\def\NAT@def@citea{\def\@citea{\NAT@separator}}
\makeatother

\begin{document}


\title{A Study on Angular Distribution of THz Radiation Driven by Two-Color Laser-induced Microplasma}

\author{
\name{Abolfazl Yousef-Zamanian and Mohammad Neshat\thanks{CONTACT M.Neshat. Email: mneshat@ut.ac.ir }}
\affil{School of Electrical and Computer Engineering, College of Engineering, University of Tehran, Tehran, Iran}
}

\maketitle

\begin{abstract}
We present an analytical method to study the angular distribution of THz radiation driven by two-color laser-induced microplasma in nitrogen gas. Directionally dependent temporal waveforms of the radiated THz pulse is calculated. To do so, we have extended the photo-current model to calculate three-dimensional distribution of the photo-induced current from which the far-field THz radiation is computed. We have also studied the effect of optical polarization state and phase difference on the angular distribution of radiation. Our study shows that for two orthogonally-polarized colors, not only THz polarization state, but also the angular distribution of THz radiation is mostly influenced by the second optical harmonic. The possibility of rotating the angular radiation pattern around the axis of the laser propagation is illustrated for circularly-polarized laser drivers by changing the optical phase difference. 

\end{abstract}


\section{Introduction}
\label{sec:1}

Terahertz (THz) radiation has been widely used in spectroscopy and imaging applications \cite{wallace2004,jepsen2011,cole2001}. For most applications in imaging, a narrow band radiation generated by e.g. photomixing, Gunn diode, or quantum cascade lasers (QCL) might be adequate \cite{liu2007,jepsen2011}. However, THz spectroscopy demands a broadband source. This is because in the entire THz spectrum, there are distinct spectroscopic fingerprints for the crystalline lattice vibrations (phonon modes), hydrogen-bonding stretches, and other intermolecular vibrations of molecules in many chemical and biological materials \cite{liu2007}.

In conventional femtosecond laser based broadband sources, phase mismatch or charge dynamics, and the damage of the active medium in high powers are hindrances to achieve high bandwidth and high intensity THz radiation. In contrast, for THz generation based on laser-induced plasma in gases, there is the possibility of continuous replacement of the gas molecules to keep the active medium safe from damage. Moreover, the absence of neither phonon bands nor boundary reflection surfaces in dry gas, and the dynamics of free electrons as charge carriers are superior advantages that result in broadband high intensity THz radiation \cite{roskos2007,Oh2013}. 

Since the introduction of photo-induced terahertz generation based on the radial acceleration of the ionized electrons \cite{hamster1993,hamster1994}, various approaches based on laser induced plasma have been proposed. DC- and AC-biased plasma methods \cite{loffler2000,loffler2002,cook2000}, few-cycle pulses without second harmonic (SH) field \cite{sheng2005}, or sub-single-cycle THz wave by three-color air plasma \cite{zhou2018}, and linear mode conversion due to tilted laser pulse incidence on the density gradient of inhomogeneous plasma \cite{sheng2005} are the most common methods to name. There have been considerable efforts to identify the physics behind these phenomena, and to find ways to increase the power of the terahertz radiation and the efficiency. In these efforts, the effect of the amplitude ratio or phase difference between two optical harmonics, length of plasma filament, frequency chirping and finding the optimal optical frequency or even the type of gases under optical irradiation have been investigated \cite{chen2015,meng2016,buccheri2018,bogatskaya2018,fedorovoptimal2018,Nguyen2018,Nguyenpra2018}. Moreover, on detection side, some groups have worked to improve air plasma-based THz detection methods \cite{nahata1998,cook1999,dai2006,wang2012}.
Recently, laser-induced microplasma (LIMP) using low energy pulse lasers has been proposed \cite{buccheri2015} which makes this technique more accessible. On the other hand, the application of terahertz radiation in studying the properties of materials in all four states of matter highlights the importance of investigating the characteristics of the generated terahertz radiation. In this regard, some papers have reported angular distribution of generated THz power or the so called radiation pattern. For example, the pattern of terahertz radiation due to micro-filament from one-color laser is shown to be perpendicular to the laser propagation axis in the studies of Buccheri et al. \cite{buccheri2015}, as well as from two-color lasers under certain conditions \cite{buccheri2018}. Smaller plasma length have forward THz power pattern, whereas longer plasma length lead to dephasing between two-color beams, and THz pattern comes in the form of a cone \cite{kumar2015,buccheri2018}. Ushakov et al. have studied the mechanism of backward terahertz radiation from laser induced microplasma \cite{ushakov2017,ushakov2019}.
It should be noted that earlier studies have been conducted on angular power distribution on different planes perpendicular to the laser propagation \cite{blank2013,ushakov2017td}, or effects of plasma density distribution and filaments length on radiation profile \cite{gorodetsky2014,you2012}.
Inspired by LIMP, in our previous study, we examined the polarization state of the generated THz wave driven by two-color compact laser-induced plasma in air, and studied the effects of the optical beams properties on the polarization state \cite{YZ2017}. That study resulted in well-balanced results with the reports of Dai \cite{dai2009} and Fedorov \cite{fedorov2016}. 

In this article, a study is presented on angular distribution of the wideband THz radiation emanating from two-color laser-induced microplasma and its dependency on optical pulse polarizations and phase difference in AC-biased approach. Moreover, directionally dependent temporal waveform of the radiated THz pulse is calculated. To do so, we have extended the photocurrent model to calculate three-dimensional distribution of the photo-induced current from which the far-field THz field pattern is obtained from electromagnetic radiation theory. It will be shown that the calculation results from the model, is in agreement with reported experimental results, as yet does not involve the complexity of other methods. In Section 2, an overview is given on the modeling of plasma-based THz generation in the AC-bias method with the emphasis on the photo-current model. The simulation results using such model for numerous cases are presented in Section 3, followed by conclusion remarks in Section 4.

\section{Modeling}
\label{sec:2}

Several models have been proposed to describe THz generation phenomenon based on compact laser-induced gas plasma. First model is a phenomenological description of this phenomenon in terms of four-wave mixing that soon became clear that it may not give the full picture \cite{kim2009}. In a second model known as AC-bias method, semi-classical equations for the optical field are used following the ionization to give a microscopic photocurrent model that accounts for electron motion \cite{kim2009}. Such model gives more insight with quantitative parameters than does the first model. The third model as the most accurate analysis of this phenomenon is through quantum theory \cite{schuh2013,karpowicz2009}. Finally, a method using ponderomotive force model to account for a nonlinear current source has been recently proposed in \cite{kumar2015}. It is worth mentioning that the intensity of the driving lasers determines the type of the dominant mechanism involved, and the best model to use \cite{berge2019}.
In this study, we use the photocurrent model described in our previous publication \cite{YZ2017}. Here for the sake of completeness, this model is briefly explained. By focusing a femtosecond laser with sufficient pulse energy on a target gas, such as nitrogen, its molecules begin to ionize. As a result, the freed electrons under the laser irradiation, start to accelerate and drift in addition to the oscillation at laser frequency. Accelerated motion of electrons emits electromagnetic wave. The dynamics of the accelerated motion (photo-current) is such that the emitted EM wave has spectrum in THz range.

Commonly the pulse energy of femtosecond lasers used for high power THz generation from air is such that based on Keldysh parameter the tunneling ionization (TI) process dominates \cite{keldysh1965}. For noble gases, the ionization rate is computed from Ammosov-Delone-Krainov (ADK) relation \cite{roskos2007}. It has been shown that for structureless atomic-like molecules such as nitrogen gas ($N_2$), ADK is a good approximation even for multiple degrees of ionization \cite{kim2012}. It should be noted that the ionization rate is related to the optical field in a highly nonlinear fashion. Such nonlinearity is the root cause of optical-to-terahertz frequency conversion. 

Free electrons absorb energy and accelerate in the optical field. However, the average electron collision time is about one picosecond that is much longer than the optical pulse duration (about a few tens of femtoseconds), therefore, the avalanche ionization does not occur under femtosecond laser excitation. Recombination of electron-ion is also negligible due to their lifetimes that are relatively long (around hundreds of picoseconds).

The dynamic of free electrons under the influence of the optical field can be modeled by the Lorentz force. It is worth noting that the ions are assumed stationary because of their heavy mass. Moreover, it is assumed that the electron velocity is far from the relativistic regime, so the force exerted by the magnetic field is negligible. Given the electron density and velocity, the transverse photo-current is calculated as in \cite{kim2012,yousef2014}. Finally, the radiated (terahertz) field from the transverse photocurrent is calculated according to the radiation theory \cite{elliott2003}.

A full detail of the calculation procedure for ionization and photocurrent generation has been already described in our previous work \cite{YZ2017}. For 3-D modeling, we used the space-time profile of a Gaussian beam representation. It is assumed that the laser-induced microplasma radiates coherently, and its interaction does not considerably affect the laser propagation \cite{thiele2017}. In order to model the optical polarization-dependent behavior, the electric field is represented in a vectorial form as \cite{YZ2017}
\begin{align}
	&\nonumber {\pmb{\xi}} = {\mathop{\rm Re}\nolimits} \left\{ \begin{array}{l}
	{{\tilde E}_x}\\
	{{\tilde E}_y}
	\end{array} \right\}\\
	& \nonumber{{\tilde E}_k} = {{\tilde E}_{k,\omega }} + {{\tilde E}_{k,2\omega }},~~k = x,y \\
	& {{\tilde E}_{k,n\omega }} = {\beta _{k,n\omega }}{e^{i{\alpha _{k,n\omega }}}}\tilde E(r,z,t,n\omega )\label{control}\\
	&\nonumber \tilde E(r,z,t,1\omega ) = {{\tilde E}_0}(r,z,t,1\omega )\\
	&\nonumber \tilde E(r,z,t,2\omega ) = \eta {e^{i\phi }}{{\tilde E}_0}(r,z,t,2\omega )
\end{align}
\noindent where ${\tilde{E}_0}(r,z,t,n\omega )$ ($n=1,2$) has the space–time profile of
a Gaussian beam as given in \cite{YZ2017}, $\eta$ is the second harmonic coefficient, $\phi$ is the phase difference (time delay) between fundamental harmonic (FH) and second harmonic (SH) lasers, and that all possible polarization states for FH and SH optical fields can be obtained by properly adjusting the parameters $\beta_{k,n\omega}$ and $\alpha_{k,n\omega}$. The effect of these parameters on the angular distribution of THz radiation will be discussed in the next Section.

The angular distribution of the radiated THz power density in frequency domain, $P_{THz}(\theta,\varphi,\omega)$, is obtained as 

\begin{align}
	&\nonumber  \ {P_{THz}}\left( {\theta ,\varphi ,\omega} \right) \propto {\left| {{\mathbf{E}_{THz}}\left( {\theta ,\varphi ,\omega} \right)} \right|^2} \\ 
	&\nonumber  \ \mathbf{E}_{THz}\left( {\theta ,\varphi ,\omega} \right)=-i\omega\mathbf{A}( {\theta ,\varphi ,\omega})\\
	&\nonumber  \ \mathbf{A} \left( {\theta ,\varphi ,\omega} \right) \propto  {\Im _\theta }\hat \theta  + {\Im _\varphi }\hat \varphi \label{Eq2} \\
	&\nonumber \ {\Im _\theta } = \iiint{\left[ \cos \theta \cos \varphi {J_x}\left( {x',y',z',\omega} \right) + \cos \theta \sin \varphi {J_y}\left( {x',y',z',\omega} \right) - \sin \theta {J_z}\left( {x',y',z',\omega} \right)\right]e^{ik\beta }}\textrm{d}x'\textrm{d}y'\textrm{d}z'\\
	& \ {\Im _\varphi } = \iiint{\left[- \sin \varphi {J_x}\left( {x',y',z',\omega} \right) + \cos \varphi {J_y}\left( {x',y',z',\omega} \right)\right]e^{ik\beta }}\textrm{d}x'\textrm{d}y'\textrm{d}z'\\
	&\nonumber \ 
	\beta  = {x'}\sin \theta \cos \varphi  + {y'}\sin \theta \sin \varphi + {z'}\cos \theta \
\end{align}

\noindent where $k=\omega/c$ and $\omega$ is the wavenumber and angular frequency at THz range, respectively, ($\theta$,$\varphi$) are the angular spherical coordinates of the observation point, $(x',y',z')$ are the source point coordinates, $\mathbf{E}_{THz}$ is the radiated THz electric field vector,  $\mathbf{A}$ is the magnetic vector potential \cite{elliott2003} and $ \ {J_{x,y,z}}\left( {x',y',z',\omega} \right)$ are the Cartesian components of the photocurrent density vector calculated from the formulations fully described in the appendices in \cite{YZ2017}. Having the field quantities in frequency domain, one can easily obtain their time-domain counterparts through inverse Fourier transform. Therefore, THz pulse energy, $(\zeta_{THz})$, at each angular direction can be calculated as

\begin{align}
\ {\zeta _{THz}}\left( {\theta ,\varphi } \right) \propto \int {{\left| {{\mathbf{E}_{THz}}\left( {\theta ,\varphi ,t} \right)} \right|^2}} \textrm{d}t \label{Eq3}
\end{align}



\section{Results and discussion}
\label{sec:3}

Using the model summarized in \Cref{sec:2}, we have already investigated the effects of laser pulse duration and intensity on the bandwidth and the peak signal of the THz point source \cite{YZ2017,yousef2014}. Moreover, using the same model, we have studied the effects of the optical polarizations on the generated THz polarization state \cite{YZ2017} that are in agreement with other reports \cite{fedorov2016}. Considering the low yield of THz power generation, knowing the spatial  distribution of the generated THz radiation becomes important so that one can adjust the optical setup parameters for the best performance. 

Here we investigate the effects of optical polarizations and phase difference on angular distribution of generated THz radiation. The spatio-temporal photocurrent distribution is first obtained by the above model, and it is transformed into frequency domain by Fourier analysis. The radiation field at each frequency with respect to the angular direction $(\theta,\varphi)$ is obtained through far field calculations from the photo-current in the frequency domain according to (\ref{Eq2}). The temporal waveforms are then calculated through inverse Fourier transformation, and finally THz pulse energy at each angular direction can be calculated from (\ref{Eq3}).

\begin{figure}[hbt!]		
	\centering
	\includegraphics[width=0.7\textwidth]{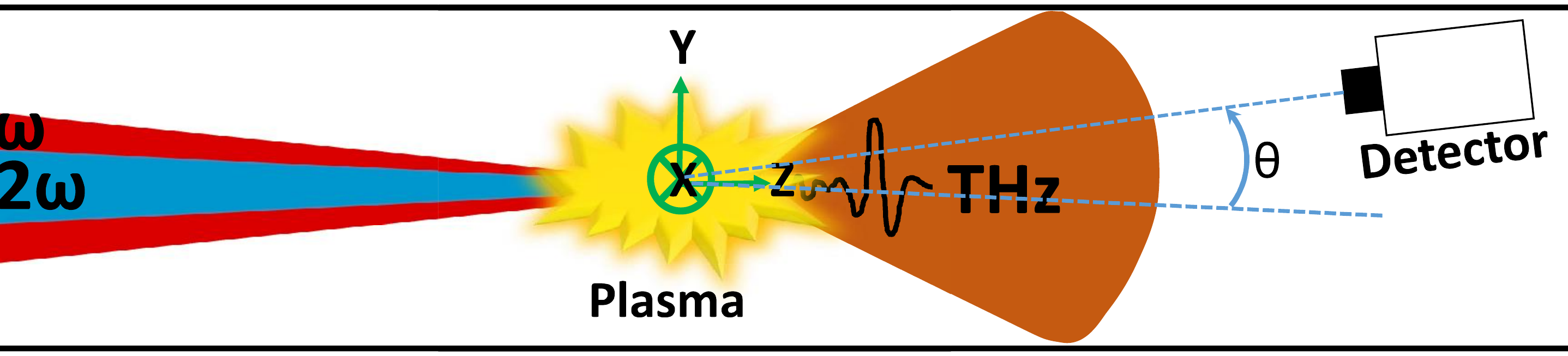}
	\vspace{-0.8\baselineskip}
	\caption{Schematic arrangement of the THz radiation driven by two-color laser-induced microplasma in air. Lasers parameters: wavelength at FH=800 nm, pulse energy $E\textrm{j}=2~\mu\textrm{J}$ ($I_0=6\times10^{14}~ \textrm{W/cm}^2$), pulse duration $\tau_{\omega,2\omega}= 25$ fs, SH intensity efficiency=$20\%$, beam width at focus $\sigma _{_0,\omega}=2~\mu \textrm{m}$, and Rayleigh range $z_R= 15.5~\mu$m. The length of  LIMP is considered to be twice of the Rayleigh range.}
	\label{fig:profile}
\end{figure}

In order to study the angular distribution, we calculated the radiation patterns and radiated THz pulse energies on $xz$-plane and $yz$-plane, and far from the microplasma spot as depicted in \Cref{fig:profile}. The plasma length for the laser characteristics given in \Cref{fig:profile} is around twice the Rayleigh range that is sufficiently smaller than dephasing length. Thus initially incident optical polarizations are assumed to be unchanged \cite{you2012}. An experimental setup has been proposed in \cite{meng2016} to provide full control over the polarization state of fundamental and second optical harmonics.

We start with a case that the optical field is a combination of an $x$-polarized Gaussian pulse and its co-polarized second harmonic. Assume the femtosecond laser has center wavelength of 800 nm, pulse energy of $E\textrm{j}=2~\mu\textrm{J}$ ($I_0=6\times10^{14}~ \textrm{W/cm}^2$), pulse duration of $\tau_{\omega,2\omega}= 25$ femtosecond, $20\%$ SH intensity efficiency, beam width of $\sigma _{_0,\omega}=2~\mu \textrm{m}$ at the focus, and both optical beams have the same Rayleigh range ($z_R=15.5~\mu$m). The length of  laser-induced microplasma in the photocurrent model simulation is considered to be twice of the Rayleigh range. It is worth mentioning that since the calculated defocusing length is almost twice of the Rayleigh range for the given laser parameters, the defocusing is not dominant  \cite{RAE1993}. Moreover, the laser power is around $2\%$ of the critical power that causes the self-focusing \cite{Liu2005}, therefore, it is assumed that the created microplasma in nitrogen gas under standard pressure has minimal effect on the laser propagation. 

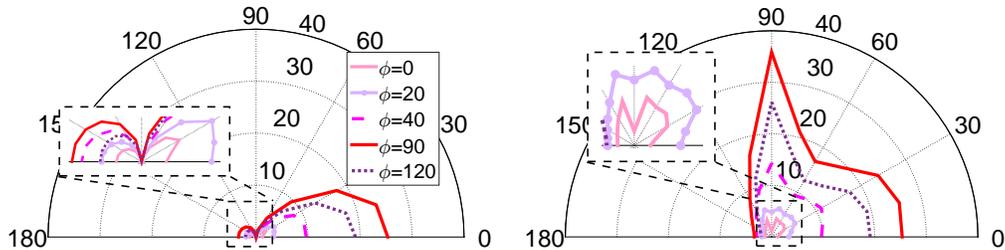
\begin{figure}[hbt!]		
	\centering	
	\begin{tikzpicture}
		\node at (0.4,-0.8) {\includegraphics[width=0.45\textwidth]{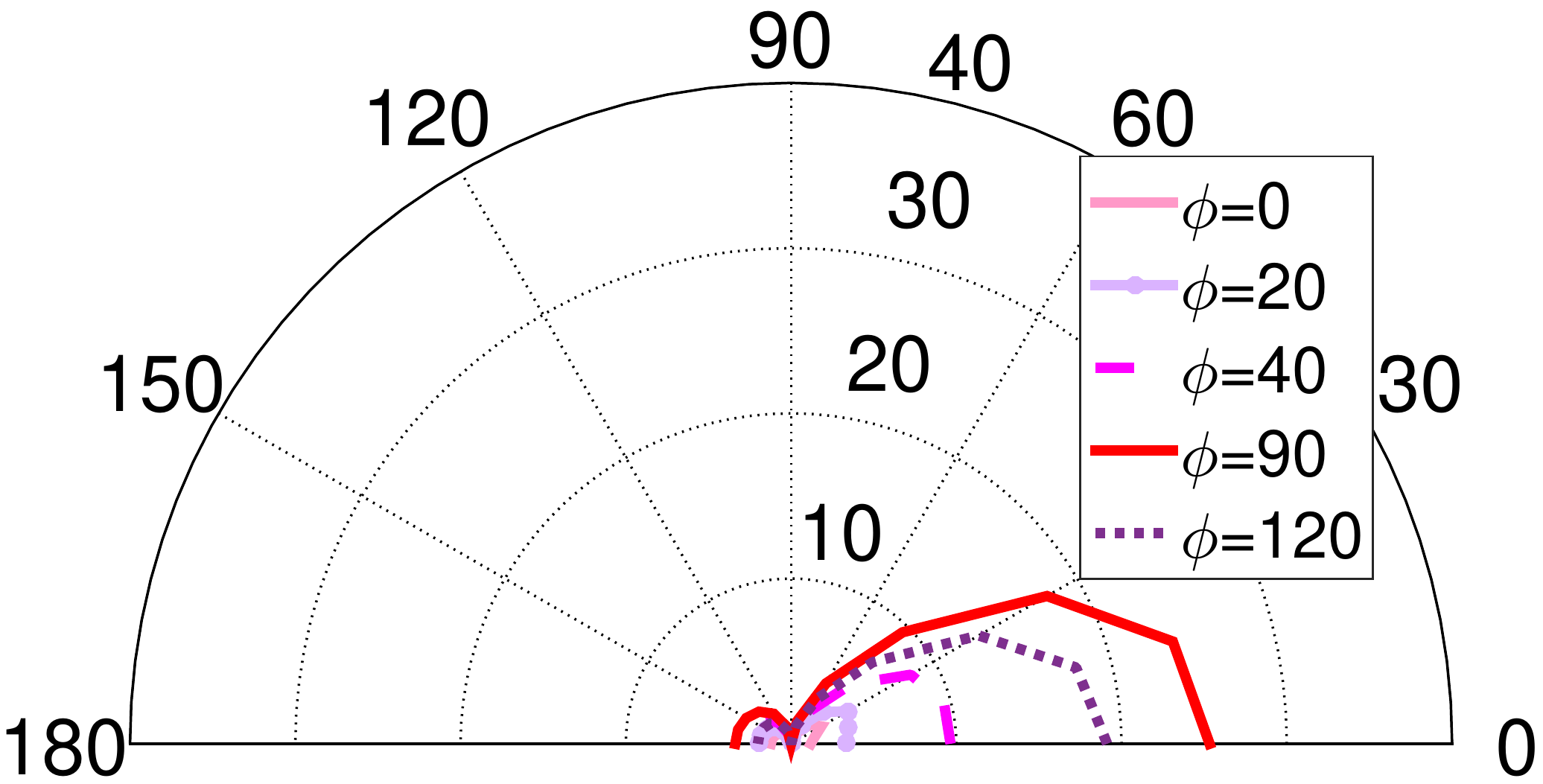}};
		\draw (-1,-1) node[draw=black,rectangle,fill=white!20,draw=black,semithick,dashed](a) {\includegraphics[width=0.14\textwidth]{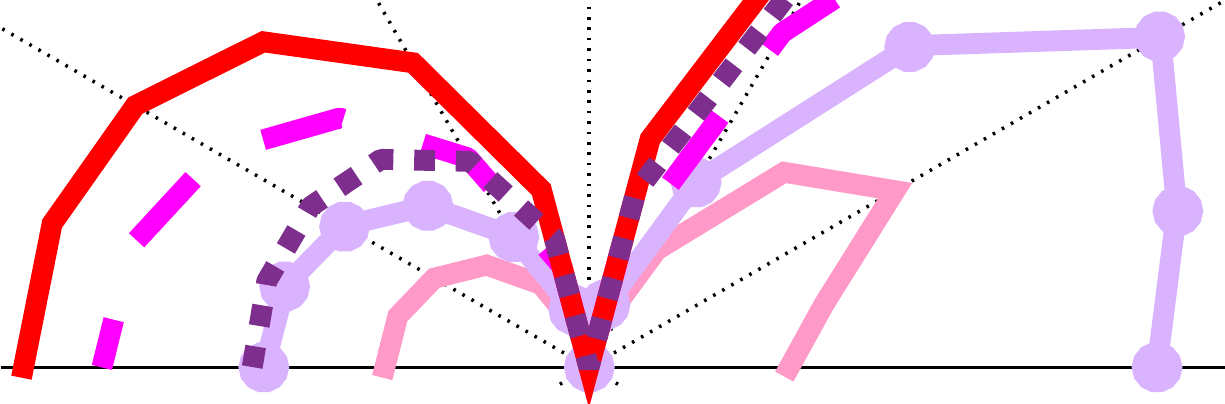}};
		\draw (0.4,-2.1) node(b)[draw=green,rectangle,draw=black,semithick,minimum size=0.6cm,dashed] {};
		\draw (a.south east)--(b.north east)[,draw=gray,rectangle,,draw=black,semithick,dashed];
		\draw (a.south west)--(b.north west)[draw=gray,rectangle,draw=black,semithick,dashed];
		\label{fig:L_L_XZ}
	\end{tikzpicture}
	\begin{tikzpicture}
		\node at (-.3,-1.25) {\includegraphics[width=0.45\textwidth]{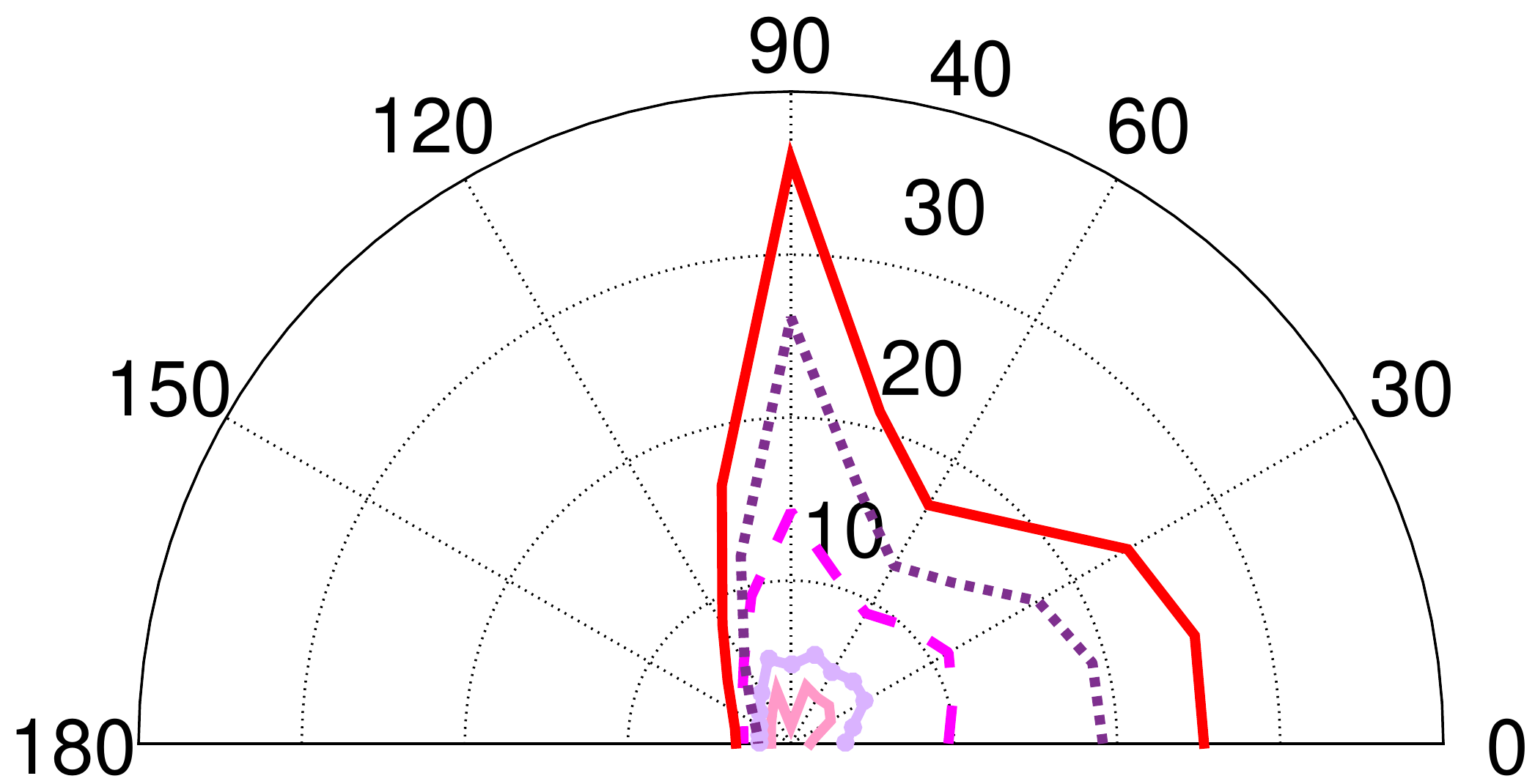}};
		\draw (-1.8,-1) node[draw=black,rectangle,fill=white!20,draw=black,semithick,dashed](a) {\includegraphics[width=0.1\textwidth]{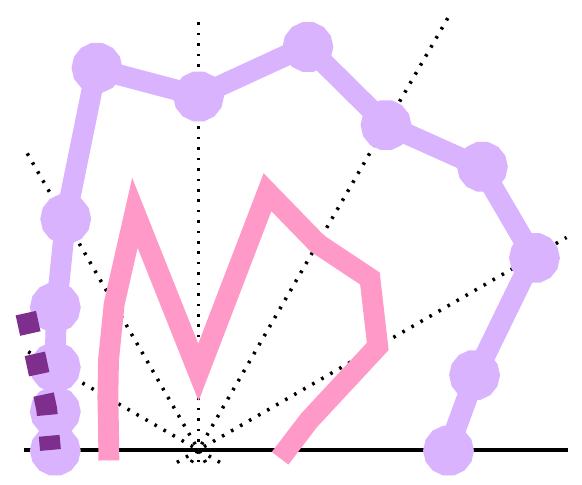}};
		\draw (-0.1,-2.55) node(b)[,draw=green,rectangle,,draw=black,semithick,minimum size=0.6cm,dashed] {};
		\draw (a.south east)--(b.north east)[,draw=gray,rectangle,,draw=black,semithick,dashed];
		\draw (a.south west)--(b.north west)[,draw=gray,rectangle,,draw=black,semithick,dashed];
	\label{fig:L_L_YZ}
	\end{tikzpicture}

	\caption{Angular distribution of THz pulse energy radiation for the $x$-copolarized lasers in $xz$-plane (left) and $yz$-plane (right). $\phi$ is the phase difference between two laser harmonics.}
	\label{fig:angular_L_L}
\end{figure}

\Cref{fig:angular_L_L} shows the angular distribution of THz pulse energy radiation for different phase differences and with the above-mentioned parameters. As expected for the $x$-copolarized condition, THz wave is maximized at $\phi=90^\circ$ \cite{YZ2017}.

Although the maximum power radiation is in $y$-direction, but the angular power distribution is narrow. In addition, when the phase difference approaches zero, forward ($z$-direction) radiation pattern becomes cone-shaped, and the backward pattern becomes comparable. Moreover, in \Cref{fig:angular_L_L}, at angle zero the THz pulse energy starts to increase from the phase of $\phi=0^\circ$ up to 90$^\circ$ and drops at 120$^\circ$. The drop in pulse energy continues up to 180$^\circ$ and this behaviour repeats itself with the periodicity of 180$^\circ$ (Not shown). Such behaviour is in agreement with the experimental results presented in Fig. 2a of \cite{meng2016}.

\begin{figure}[hbt!]	
	\centering
	{\includegraphics[width=0.8\textwidth]{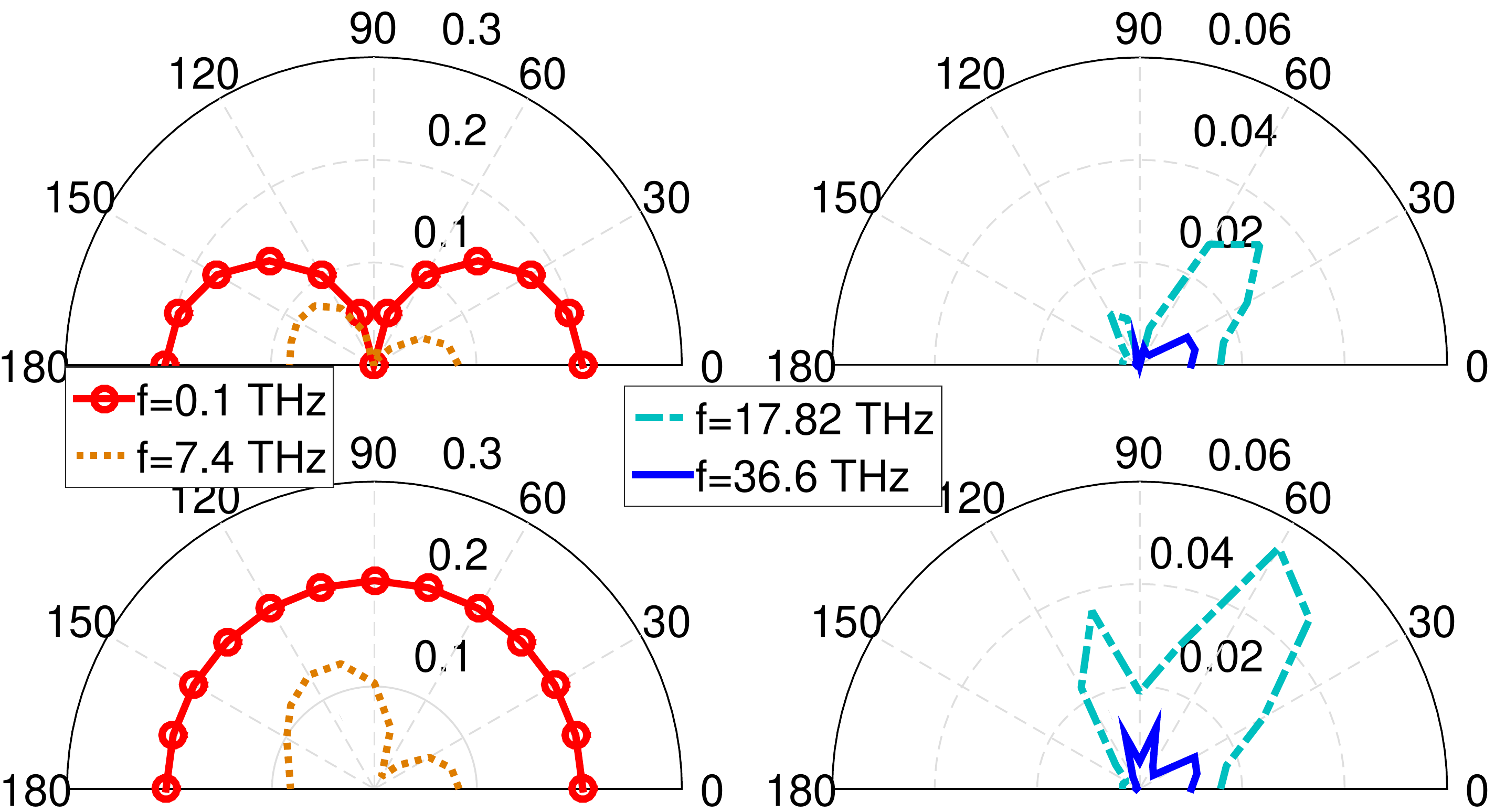}
	\caption{Radiation pattern of THz waves at different frequencies for $\phi=0^\circ$ in $xz$-plane (top) and $yz$-plane (bottom).}
	\label{fig:radiation_pattern_L_L_P_0}}	
	{\includegraphics[width=0.8\textwidth]{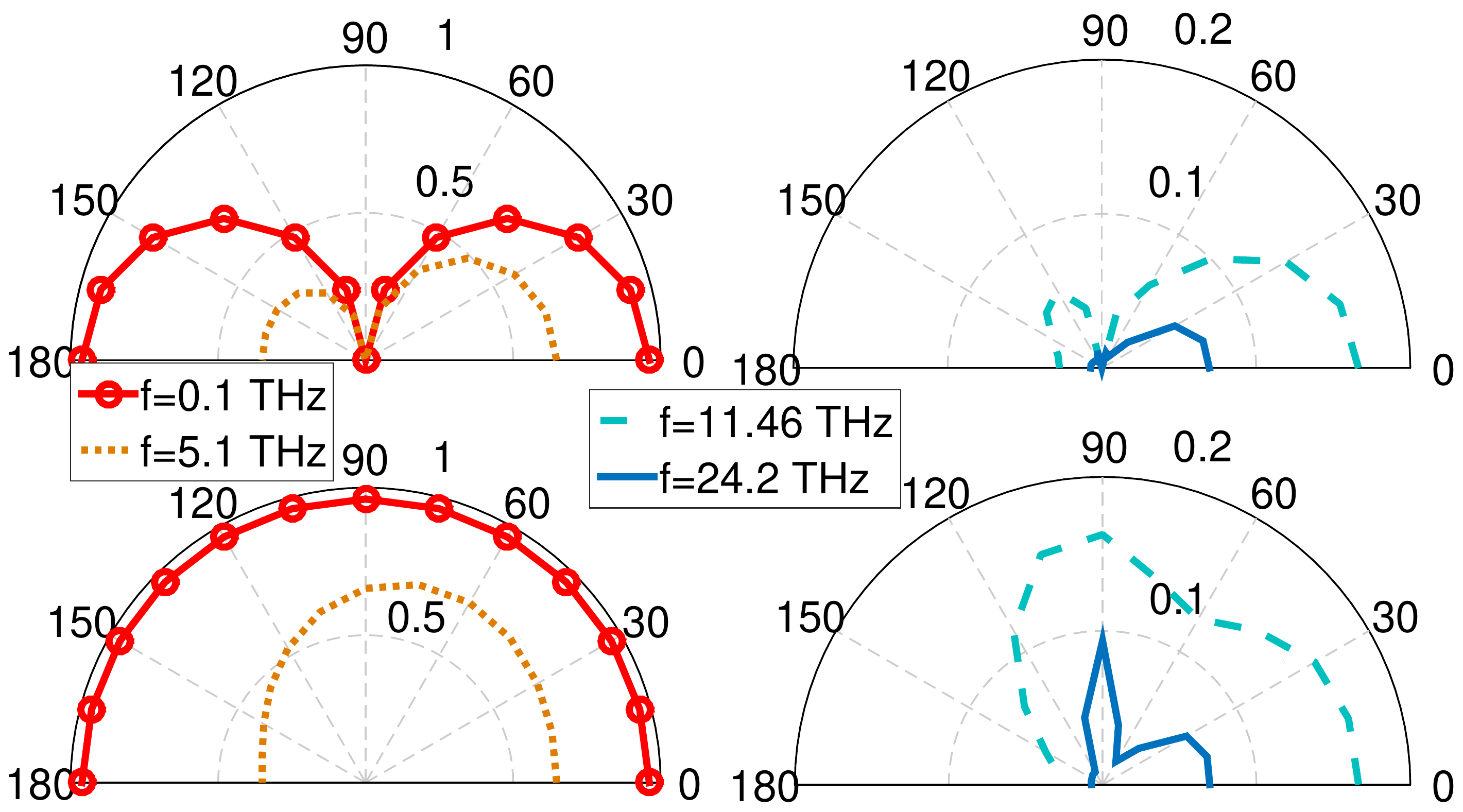}}
	\caption{Radiation pattern of THz waves at different frequencies for $\phi=90^\circ$ in $xz$-plane (top) and $yz$-plane (bottom).}
	\label{fig:radiation_pattern_L_L_P_90}
\end{figure}

\Cref{fig:radiation_pattern_L_L_P_0,fig:radiation_pattern_L_L_P_90} show the THz radiation pattern behavior with respect to the frequency for $\phi=0^\circ$ (or $180^\circ$) and $\phi=90^\circ$, respectively. The radiation patterns are calculated in frequency domain according to (\ref{Eq2}). The frequencies are chosen such that the difference between their corresponding radiation patterns becomes appreciable. It is evident that lower frequencies have dipole-like radiation pattern, i.e. omni-directional in the plane perpendicular to the current dipole ($yz$-plane) and donut-shaped in the plane of dipole ($xz$-plane) \cite{elliott2003}. At higher frequencies, the radiation pattern becomes more complex in shape as the spatially distributed photocurrent may become electrically large as compared to the radiation wavelength. Therefore, the radiation in each direction is obtained by summing over all small segments of the photocurrent (or dipoles) at far-field according to the integrals given in (\ref{Eq2}).

\begin{figure}[hbt!]
	\begin{tikzpicture}
		\node at (-.3,-1.25) {\includegraphics[width=0.45\textwidth]{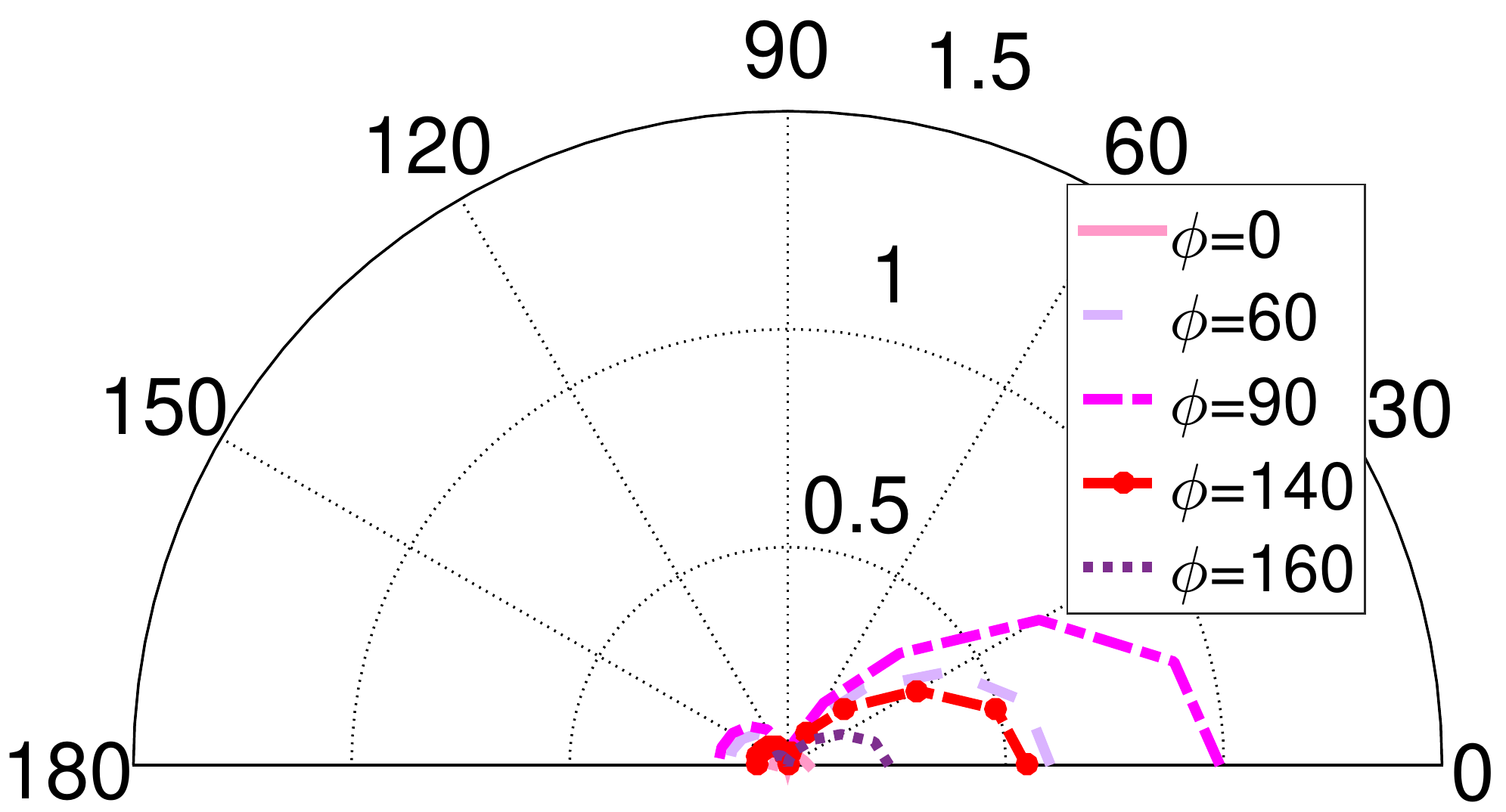}};
		\draw (-1.7,-1) node[draw=black,rectangle,fill=white!20,draw=black,semithick,dashed](a) {\includegraphics[width=0.14\textwidth]{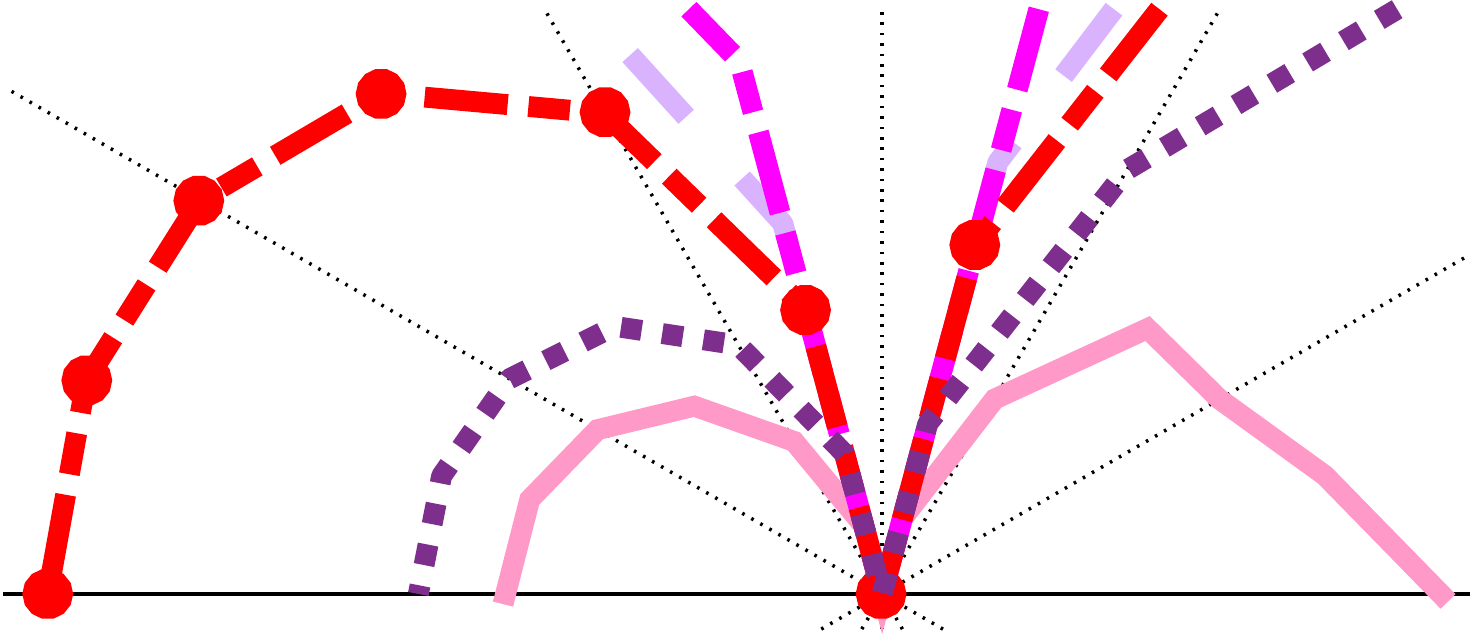}};
		\draw (-0.3,-2.6) node(b)[,draw=green,rectangle,,draw=black,semithick,minimum size=0.65cm,dashed] {};
		\draw (a.south east)--(b.north east)[,draw=gray,rectangle,,draw=black,semithick,dashed];
			\draw (a.south west)--(b.north west)[,draw=gray,rectangle,,draw=black,semithick,dashed];
		\label{fig:L+L_XZ}
	\end{tikzpicture}
	\begin{tikzpicture}
		\node at (-.3,-1.23) {\includegraphics[width=0.45\textwidth]{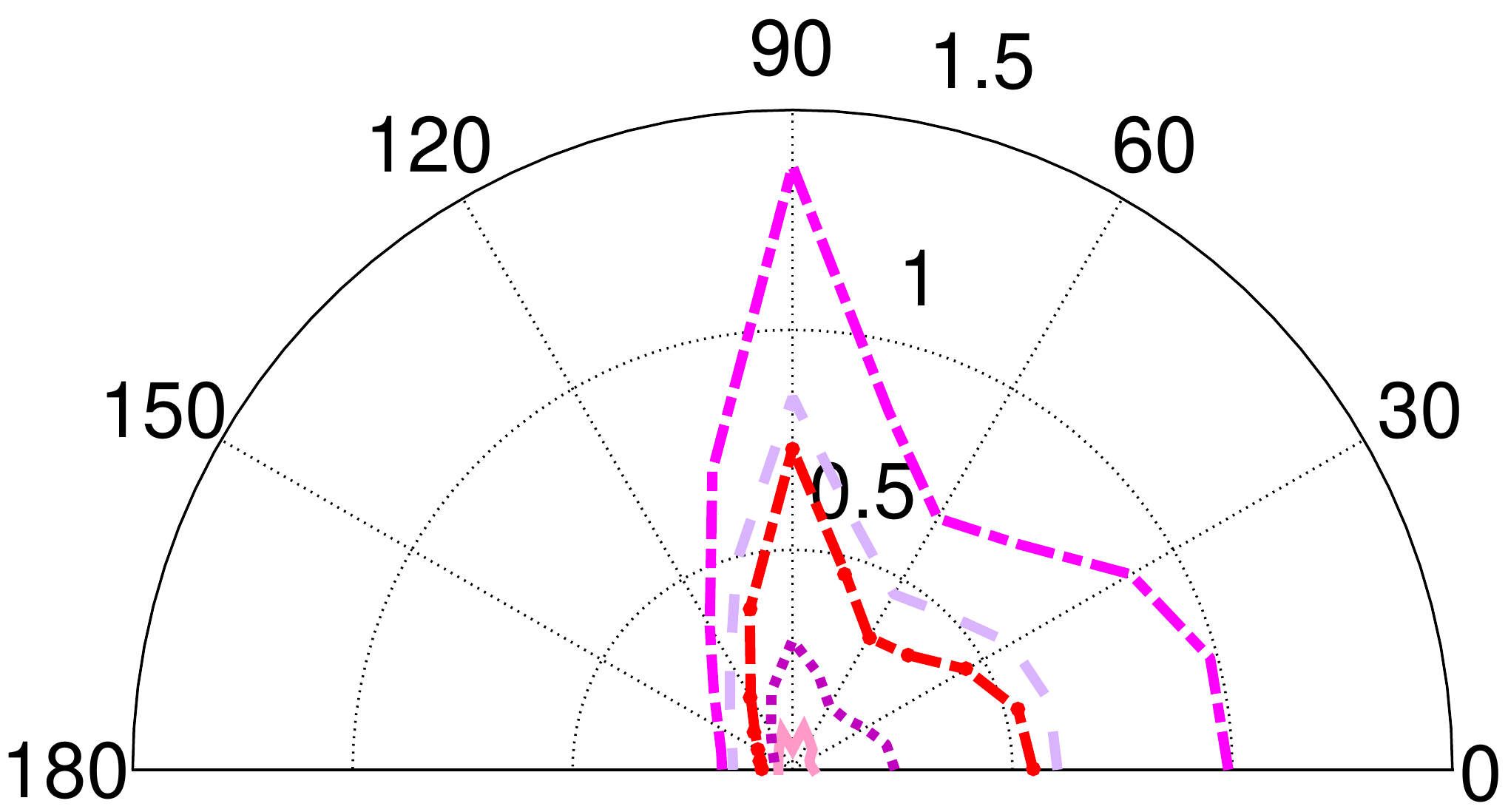}};
		\draw (-1.7,-1) node[draw=black,rectangle,fill=white!20,draw=black,semithick,dashed](a) {\includegraphics[width=0.1\textwidth]{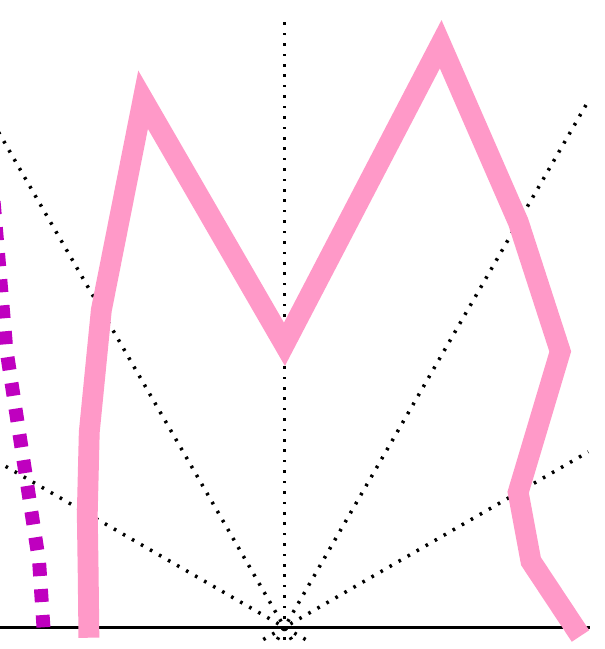}};
		\draw (-0.2,-2.6) node(b)[,draw=green,rectangle,,draw=black,semithick,minimum size=0.65cm,dashed] {};
		\draw (a.south east)--(b.north east)[,draw=gray,rectangle,,draw=black,semithick,dashed];
		\draw (a.south west)--(b.north west)[,draw=gray,rectangle,,draw=black,semithick,dashed];
		\label{fig:L+L_YZ}
	\end{tikzpicture}

	\caption{Angular distribution of THz pulse energy for the linear and orthogonal optical polarizations in $xz$-plane (left) and $yz$-plane (right).}
	\label{fig:angular_L+L}
\end{figure}
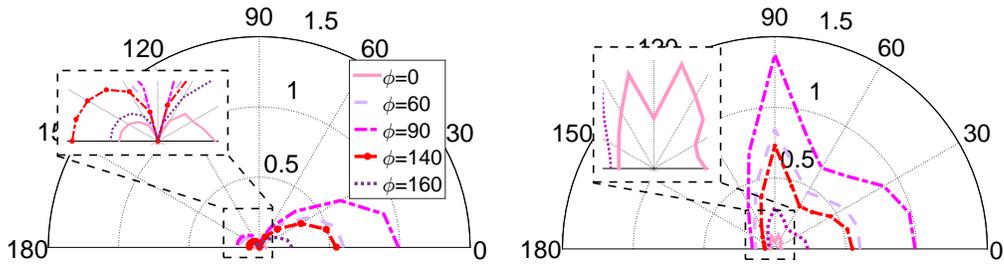

\Cref{fig:angular_L+L} shows the angular distribution of THz pulse energy for the case that fundamental and second harmonic optical polarization states are linear but orthogonal along $y$- and $x$-direction, respectively, whereas other parameters are kept exactly the same as before. It is worth noting that the angular distribution, as can be seen in \Cref{fig:angular_L+L}, is mostly influenced by the second optical harmonic. It has been observed that the terahertz wave polarization is also influenced mostly by the second optical harmonic \cite{YZ2017,kumar2015}, as if the fundamental harmonic predominantly contributes to ionization, whereas the second harmonic mainly determines the characteristics of THz generation.

\begin{figure}[hbt!]		
	\includegraphics[width=0.47\textwidth]{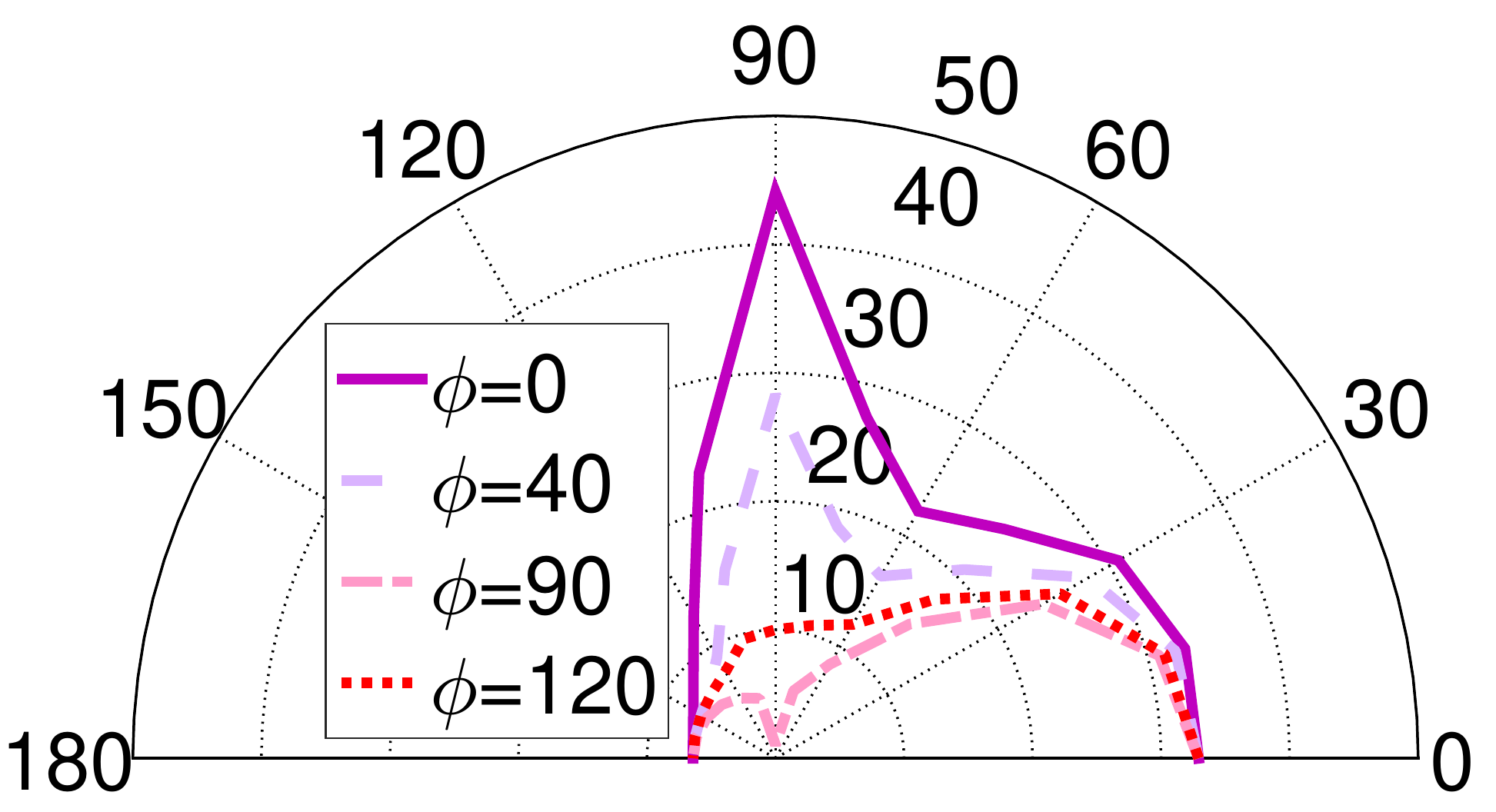} \label{fig:CC_XZ}
	\includegraphics[width=0.47\textwidth]{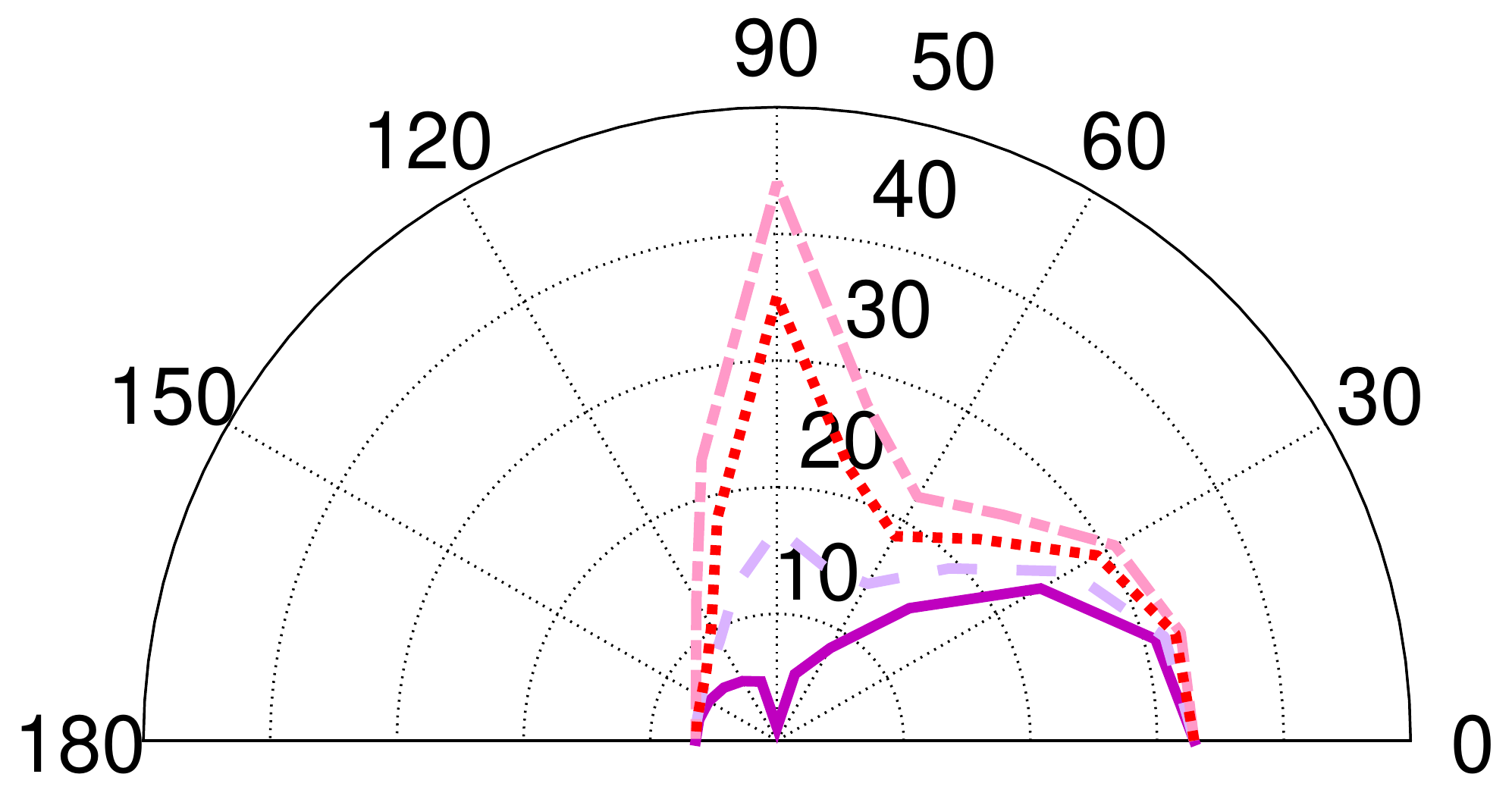} \label{fig:CC_YZ}
	\caption{Angular distribution of THz pulse energy for the left handed circular polarizations in $xz$-plane (left) and $yz$-plane (right). }
	\label{fig:angular_CC}	
\end{figure}

In the last case, both optical beams have left handed circular polarization states, where the other parameters are kept  exactly the same as before. As seen in \Cref{fig:angular_CC}, the radiation pattern of THz waves rotates around the laser propagation axes when the phase difference between two harmonics increases. This phenomenon can have interesting practical implications, e.g. THz beam steering. Moreover, comparing \Cref{fig:angular_CC} and \Cref{fig:angular_L_L}, it is evident that higher THz pulse energy emission occurs for circularly polarized as compared to linearly polarized laser driver. Therefore, circularly polarized laser driver leads to higher laser-to-THz conversion efficiency as also reported in \cite{Tailliez2020}. At angle of zero in \Cref{fig:angular_CC}, the THz pulse energy stays almost constant by changing the phase from $\phi=0^\circ$  to 120$^\circ$ (our calculation continued up to 370$^\circ$ showing the constant pulse energy but not shown). Such behaviour is also in agreement with the experimental results presented in Fig. 2a of \cite{meng2016}.

\begin{figure}[hbt!]
	\centering
	\includegraphics[width=0.47\textwidth]{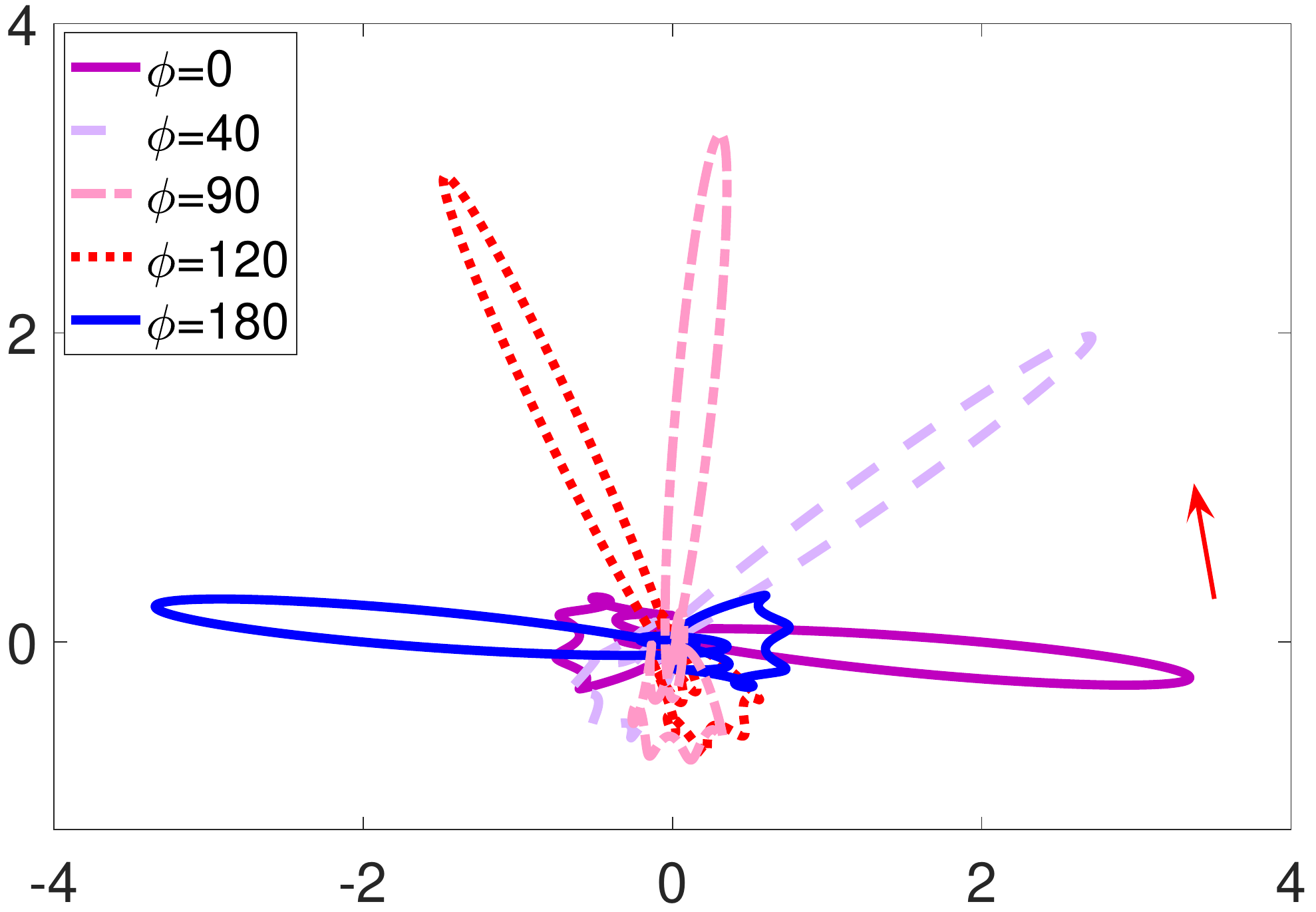} 
	\caption{Time evolution of THz electric field vector in principle polarization coordinate with left handed circular polarization for both lasers.}
	\label{fig:CC_pol_jom}
\end{figure}

As seen in \Cref{fig:angular_CC,fig:CC_pol_jom}, the advantage of this case is that by changing the phase difference, amplitude of the THz field does not drop, and the radiation pattern rotates around the laser propagation axis. Moreover, as seen in \Cref{fig:CC_pol_jom}, the polarization state of the terahertz wave is approximately linear, and its tilt angle rotates counter clockwise by increasing the phase difference between fundamental and second optical harmonics. This behavior can be used in the non-destructive investigation of the properties of materials exposed to THz wave.

\begin{figure}[hbt!]		
	\includegraphics[width=0.47\textwidth]{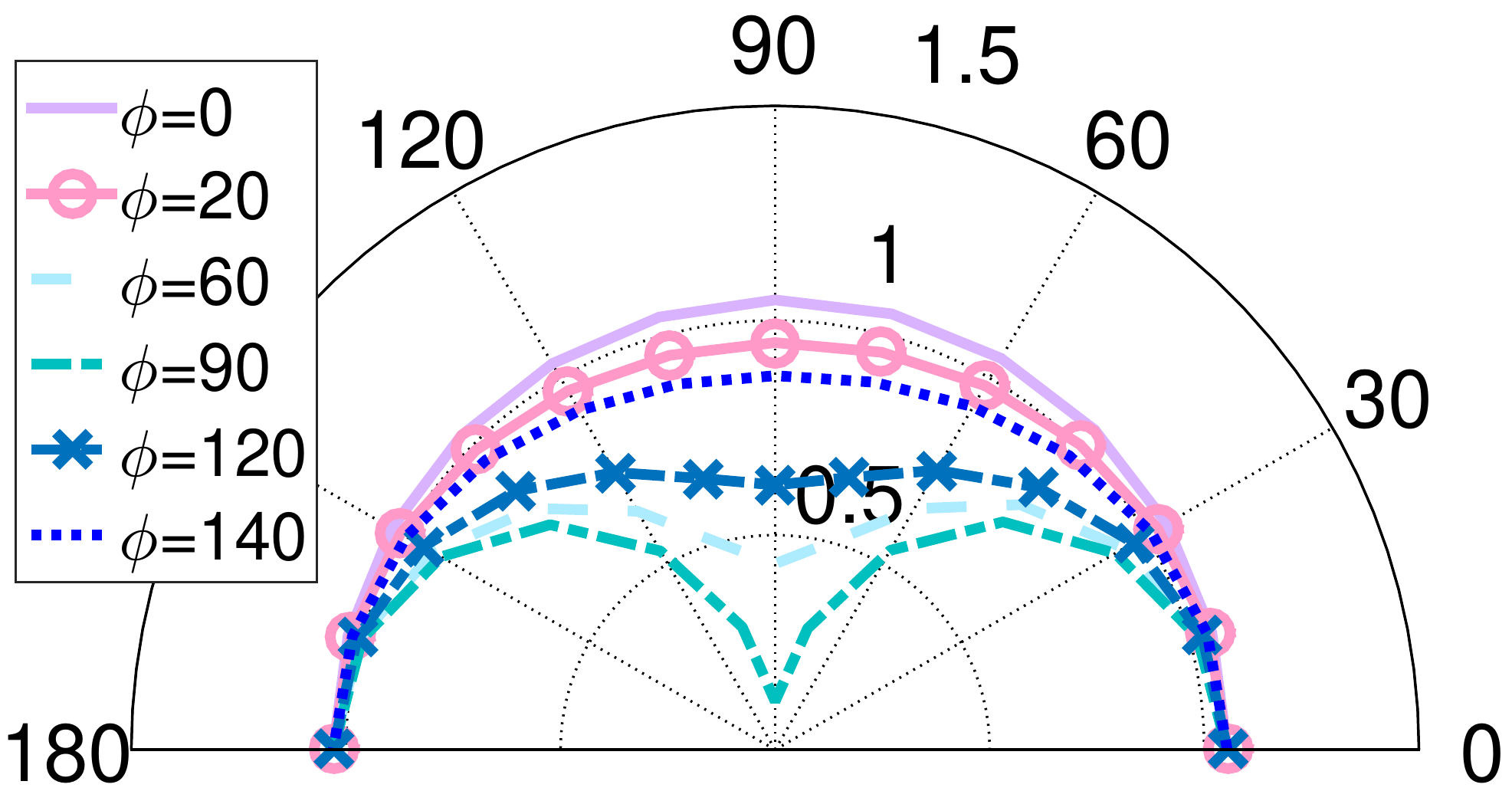} \label{fig:CC_XZ_Freq_phi}
	\includegraphics[width=0.47\textwidth]{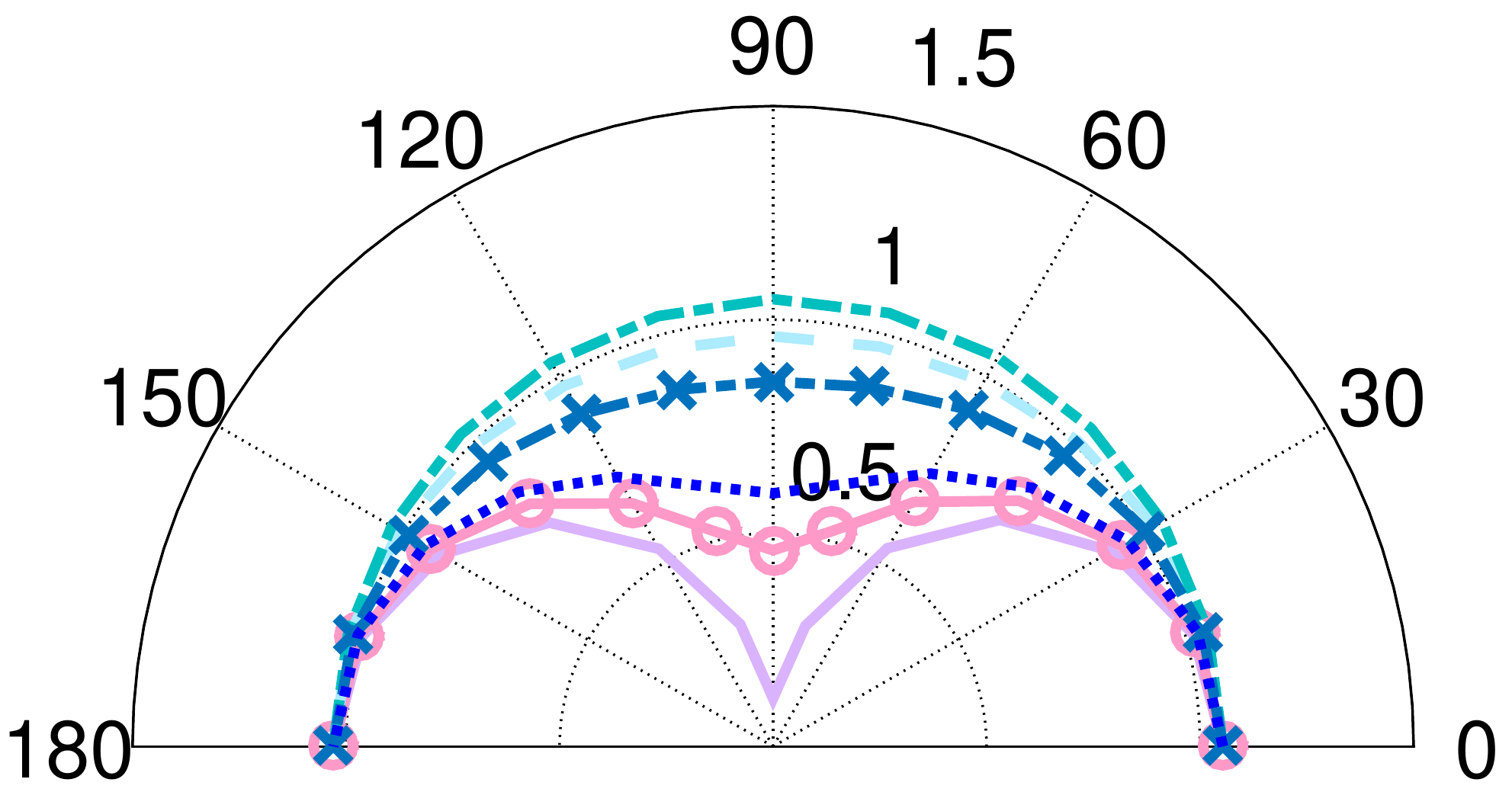} \label{fig:CC_YZ_Freq_phi}
	\caption{Radiation pattern of THz wave at $1.3$ THz in $xz$-plane (left) and $yz$-plane (right).}
	\label{fig:radiation_pattern_CC_phi}
\end{figure}

\begin{figure}[hbt!]		
	\centering
	\includegraphics[width=0.8\textwidth]{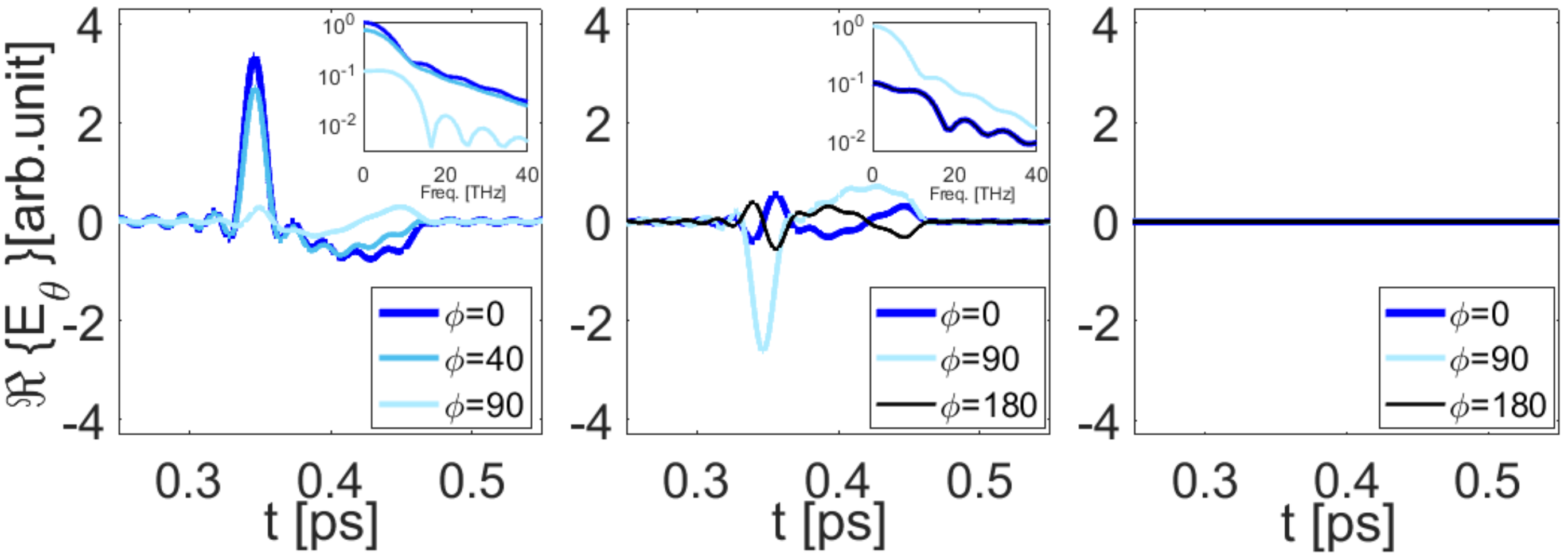} \label{fig:CC_Re_ETH}
	\includegraphics[width=0.8\textwidth]{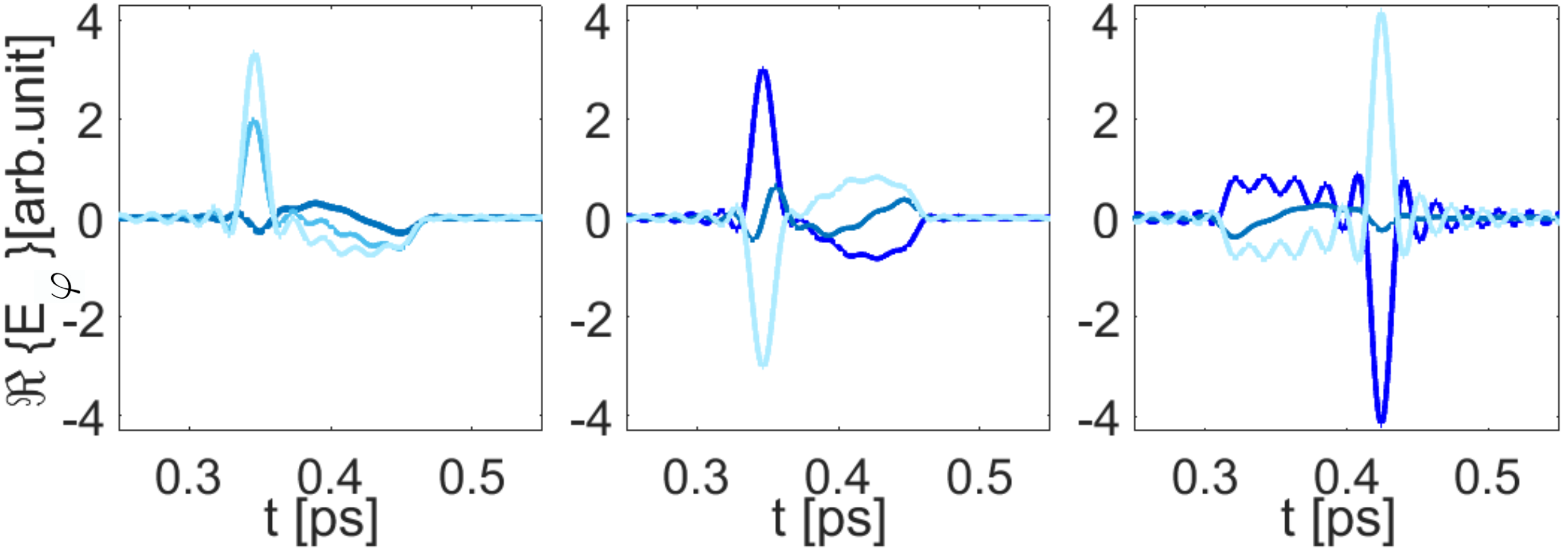} \label{fig:CC_Re_EPH}
	\caption{Calculated temporal waveforms of THz pulse components of $E_\theta$(top) and  $E_\varphi$(bottom) along $\theta=0^\circ$ (left), $\theta=30^\circ$ (middle), and $\theta=90^\circ$ (right). Insets show the corresponding frequency spectrum of THz pulses.}
	\label{fig:Temporal_CC}
\end{figure}

\Cref{fig:Temporal_CC} shows the calculated temporal wave forms of the generated THz field components of $E_\theta$ and  $E_\phi$ along directions of $\theta=0^\circ, 30^\circ,$ and $90^\circ$ ($\theta$ is shown in \Cref{fig:profile}). Insets shown in \Cref{fig:Temporal_CC} illustrate the corresponding frequency spectrum of THz pulses. As expected from the radiation theory, in \Cref{fig:Temporal_CC}, $E_\theta$ component of electric field along $\theta=90^\circ$ direction is zero since the driving photocurrent density vector is transverse to this direction.

\section{Conclusion}
\label{sec:4}

We computed angular distribution of the radiated THz pulse energy, and radiation pattern at different frequencies driven by two-color laser-induced microplasma as THz source. For linearly co-polarized  $\omega - 2\omega$ optical pulses, phase difference between $\omega - 2\omega$ dominantly affects the angular distribution of THz pulse energy. Accordingly, radiation pattern changes from that of dipole-like current at lower frequencies to that of spatially distributed current at higher frequencies. For linearly perpendicular optical polarizations, the same behavior is observed but with a much lower amplitude, and it is mostly influenced by the second optical harmonic. Furthermore, the study of left handed circularly polarized  $\omega - 2\omega$ optical pulses showed that by changing the optical phase difference, the amplitude of the THz field does not drop, and the radiation pattern and consequently angular THz pulse energy distribution rotates around the laser propagation axis. This phenomenon can have interesting practical implications, e.g. on THz beam steering. Finally, temporal waveform of THz pulses is shown to be direction-dependent. 

\section*{Acknowledgments}

The authors would like to acknowledge the financial support from Iran National Science Foundation (INSF) for this project.

\section*{Disclosure statement}

No potential conflict of interest was reported by the author(s).









\bibliographystyle{tfnlm}
\bibliography{Pattern_LIMP}

\end{document}